\documentclass[%
 reprint,
 superscriptaddress,
 amsmath,amssymb,
 aps,
floatfix,
]{revtex4-2}

\newcommand{\eref}[1]{Eq.~(\ref{#1})}
\newcommand{\tref}[1]{Table.~\ref{#1}}
\newcommand{\fref}[1]{Fig.~\ref{#1}}

\newcommand{\panel}[1]{(#1)}
\newcommand{\cpanel}[1]{\textbf{(#1)}}

\newcommand{\purityeq}{$(\langle \sigma_x \rangle^2 + \langle \sigma_y \rangle^2 + \langle \sigma_z \rangle^2)/2 + 1/2$}
\newcommand{\approxpurity}{$\langle \sigma_z \rangle^2/2 + 1/2$}

\usepackage{graphicx}
\usepackage{dcolumn}
\usepackage{bm}
\usepackage[colorlinks]{hyperref}

\hypersetup{%
    plainpages=true,
    breaklinks=true,
    hypertexnames=false,
    pageanchor=true,
    colorlinks=true,
    linkcolor={blue},
    citecolor={blue},
    urlcolor={blue},
    anchorcolor={black}
}
\usepackage{hhline}

\usepackage{braket, siunitx}

\begin{document}

\preprint{APS/123-QED}
\preprint{MIT-CTP/5757}

\title{Qubit-State Purity Oscillations from Anisotropic Transverse Noise}

\def\RLEaffil{Research Laboratory of Electronics, Massachusetts Institute of Technology, Cambridge, MA 02139, USA}
\def\LLaffil{Lincoln Laboratory, Massachusetts Institute of Technology, Lexington, MA 02421, USA}
\def\Physaffil{Department of Physics, Massachusetts Institute of Technology, Cambridge, MA 02139, USA}
\def\EECSaffil{Department of Electrical Engineering and Computer Science, Massachusetts Institute of Technology, Cambridge, MA 02139, USA}
\def\affilAQ{\textit{Atlantic Quantum, Cambridge, MA 02139}}
\def\affilHRL{\textit{HRL Laboratories, Malibu, CA 90265}}
\def\affilGoogle{\textit{Google Quantum AI, Goleta, CA 93111}}
\def\Tokyoaffil{Department of Applied Physics, Graduate School of Engineering, The University of Tokyo, Bunkyo-ku, Tokyo 113-8656, Japan}

\author{David~A.~Rower}
\email[These authors contributed equally;~]{rower@mit.edu}
\affiliation{\Physaffil}
\affiliation{\RLEaffil}

\author{Kotaro~Hida}
\email[These authors contributed equally;~]{hida@qipe.t.u-tokyo.ac.jp}
\affiliation{\Tokyoaffil}

\author{Lamia~Ateshian}
\affiliation{\RLEaffil}
\affiliation{\EECSaffil}

\author{Helin~Zhang}
\affiliation{\RLEaffil}

\author{Junyoung~An}
\affiliation{\RLEaffil}
\affiliation{\EECSaffil}

\author{Max~Hays}
\affiliation{\RLEaffil}

\author{Sarah~E.~Muschinske}
\affiliation{\RLEaffil}
\affiliation{\EECSaffil}

\author{Christopher~M.~McNally}
\affiliation{\RLEaffil}
\affiliation{\EECSaffil}

\author{Samuel~C.~Alipour-Fard}
\affiliation{Center for Theoretical Physics, Massachusetts Institute of Technology, Cambridge, MA 02139, USA}
\affiliation{The NSF AI Institute for Artificial Intelligence and Fundamental Interactions}

\author{R\'eouven Assouly }
\affiliation{\RLEaffil}

\author{Ilan T. Rosen}
\affiliation{\RLEaffil}

\author{Bethany~M.~Niedzielski} 
\affiliation{\LLaffil} 

\author{Mollie~E.~Schwartz}
\affiliation{\LLaffil}

\author{Kyle~Serniak} 
\affiliation{\RLEaffil}
\affiliation{\LLaffil} 

\author{Jeffrey~A.~Grover}
\affiliation{\RLEaffil}


\author{William~D.~Oliver}
\email{william.oliver@mit.edu}
\affiliation{\Physaffil} 
\affiliation{\RLEaffil} 
\affiliation{\EECSaffil} 

\date{\today}

\begin{abstract}
We explore the dynamics of qubit-state purity in the presence of transverse noise that is anisotropically distributed in the Bloch-sphere XY plane. We perform Ramsey experiments with noise injected along a fixed laboratory-frame axis and observe oscillations in the purity at twice the qubit frequency arising from the intrinsic qubit Larmor precession. We probe the oscillation dependence on the noise anisotropy, orientation, and power spectral density, using a low-frequency fluxonium qubit. Our results elucidate the impact of transverse noise anisotropy on qubit decoherence and may be useful to disentangle charge and flux noise in superconducting quantum circuits.
\end{abstract}

\maketitle

Quantum systems experience decoherence in the presence of uncontrolled environmental degrees of freedom. When engineering quantum systems for applications such as computing~\cite{Bluvstein2024, Acharya2024} or metrology~\cite{Degen2017}, decoherence poses a significant challenge to conducting high-precision experiments.
Consequently, understanding the structure of coherence-limiting noise and corresponding decoherence dynamics is of both high interest and value. Although significant effort has been put into the development of qubit decoherence models~\cite{Wangsness1953, Redfield1957, Leggett1987, Breuer2002} and noise spectral estimation~\cite{bylander2011noise, Yan2012, smith2019design, Sung2019}, signatures of the anisotropy of transverse noise (i.e., noise that is not isotropically distributed in the Bloch-sphere XY plane) in qubit decoherence~\cite{slichterPrinciplesMagneticResonance1990, vacchini_advances_2019, Choi2022, Chen2022} remains relatively underexplored~\footnote{Although such noise is a feature of the extensively-studied spin-boson model in the regime of zero bias~\cite{Leggett1987, Sun2024} (see, e.g., Section IIIB of Ref.~\cite{Leggett1987}), prior work primarily focused on dynamics of the spin polarization along one axis rather than that of the state purity. In the weak-coupling limit, the polarization dynamics are those of the damped harmonic oscillator (see Eq. 3.10 of Ref.~\cite{Leggett1987}), displaying no features at twice the qubit frequency. By studying state purity, we elucidate dynamics related to decoherence, unobscured by Larmor motion.}, with the notable exception of squeezed light inducing anisotropic rotating-frame transverse relaxation times~\cite{Gardiner1986,Murch2013, Toyli2016}. 
These prior studies focused on squeezed light in a narrow bandwidth around the qubit frequency, resulting in dynamics consistent with the rotating-wave approximation (RWA), given by the optical Gardiner-Bloch equations~\cite{Gardiner1985,Gardiner1986}.

We instead ask the following question: when exposed to large-bandwidth, off-resonant noise resulting in physics beyond the RWA~\cite{Zeng2012, Makela2013}, are there distinct signatures of transverse noise anisotropy in qubit decoherence?
Such signatures may inform a method to disentangle noise sources coupled to distinct transverse axes~\cite{Hendrickx2024}, e.g., charge and flux noise in circuit quantum electrodynamics. 
We focus on classical rather than quantum noise~\cite{clerk2010introduction}, and utilize linear polarization of the noise (rather than squeezing) to probe transverse noise anisotropy. 
Our approach, leveraging injected classical noise~\cite{Gneiting2016,Gneiting2020}, bypasses the bandwidth limitations and experimental complexity of engineered squeezed light sources~\cite{Andersen2016, Qiu2023}.

\begin{figure*}[!htb]
\includegraphics[width=\textwidth]{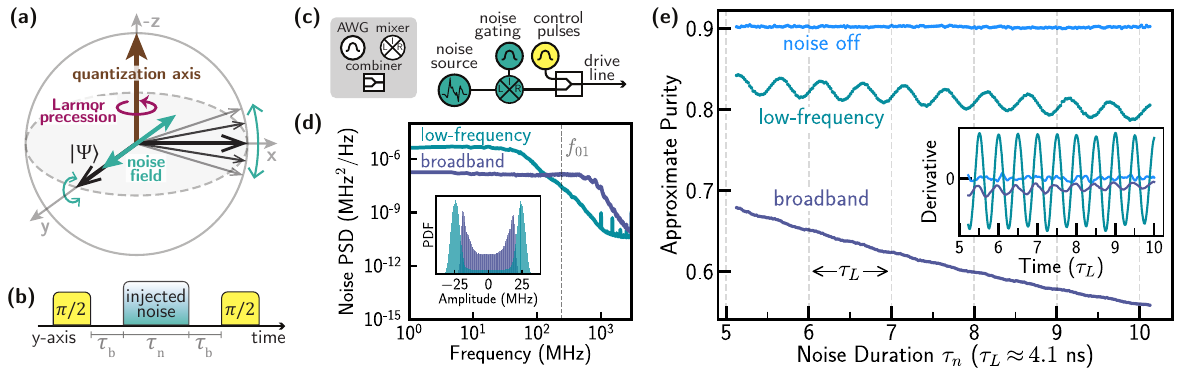}
\caption{
\label{fig1}
\textbf{Purity oscillations in Ramsey experiments with anisotropic noise.}
\cpanel{a} Bloch-sphere representation of a qubit in the lab frame undergoing Larmor precession (purple) under the influence of linearly polarized transverse noise (turquoise). 
When the qubit state is aligned with (perpendicular to) the noise axis, the state is insensitive (sensitive) to the noise, resulting in purity oscillations at twice the qubit frequency due to the Larmor precession.
\cpanel{b} Pulse sequence.
A superposition state is prepared, and noise is turned on after a buffer time $\tau_b$.
The noise is kept on for a duration $\tau_n$.
\cpanel{c} Experimental setup for noise injection.
Noise is generated, gated with a mixer, and combined with coherent control pulses on the device charge or flux line.
\cpanel{d} Injected noise power spectral densities for two noise configurations: (1) low-frequency white noise up to $\SI{100}{MHz} < f_{01}$, and (2) broadband white noise up to $\SI{1}{GHz} > f_{01}$. Inset: noise amplitude distributions. We attribute the bimodal nature of the noise to saturation of the mixer used for noise gating.
\cpanel{e} Approximate purity, $\gamma_\text{approx} \equiv \langle \sigma_z \rangle^2/2 + 1/2$, for Ramsey experiments with injected charge noise ($\hat{n} \propto \sigma_y$) in three configurations: noise off (light blue), white noise up to $\SI{100}{MHz} < f_{01}$ (turquoise), and white noise up to $\SI{1}{GHz} > f_{01}$ (dark blue). Insert: time derivatives of the approximate purity data, smoothed with a triangular window function of size $0.22\tau_L$ to clarify the signal.
}
\end{figure*}

We now introduce the central theoretical idea of this work [\fref{fig1}\panel{a}]. Consider a qubit described by density matrix $\rho$, which experiences noise that is coupled to a transverse axis in the lab frame (e.g. $H_\text{noise}/\hbar = \eta_{y}(t)\sigma_{y}$ where $\eta_{y}(t)$ is real-valued and referred to as linearly polarized, since the noise field couples to a single Bloch-sphere axis). Starting in a superposition state, the state purity $\text{tr}\left(\rho^2\right)$---indicating the degree of decoherence or disorder---will display an oscillatory component at twice the qubit frequency. We can understand this behavior by considering the qubit Larmor precession: when aligned with (perpendicular to) the noise axis, the qubit state will be insensitive (sensitive) to rotations generated by the noise field. Consequently, as the qubit precesses, the susceptibility of the state to the noise will oscillate at twice the qubit frequency.
Qubit purity oscillations~\cite{Gneiting2020, Chen2022} do not typically appear in conventional descriptions of decoherence, such as through isotropic transverse noise~\cite{slichterPrinciplesMagneticResonance1990} or coupling to a (squeezed) electromagnetic field within the RWA~\cite{vacchini_advances_2019,Gardiner1986,Murch2013}.
However, such oscillations should arise from noise in transversely-coupled qubit-control electronics, such as those used for charge or flux drives in superconducting circuits~\cite{Krantz2019}, or for laser-amplitude modulation in neutral atom experiments~\cite{Levine2022}.

In this Letter, we directly measure time-domain purity oscillations in the state of a superconducting fluxonium qubit via injected-noise experiments. We establish that anisotropic transverse noise generates periodic behavior in qubit purity. For noise with a short correlation time (Markovian noise), we observe a step-wise monotonic decay in purity, whereas longer noise correlation times result in purity revivals. We probe the noise-axis dependence of the oscillations, finding a period-doubling behavior for noise approaching the quasistatic limit (where noise is constant during a single experiment, but changes shot-to-shot). We also provide analytical calculations for the purity decay of a qubit subject to noise in the quasistatic and Markovian limits.

Our experimental platform comprises a superconducting fluxonium qubit~\cite{Manucharyan2009} of transition frequency $f_{01}=\omega_{01}/2\pi\approx\SI{243.7}{MHz}$ (Larmor period $\tau_L = f_{01}^{-1} \approx \SI{4.1}{ns}$) with individual charge and flux control, cooled to $\approx\SI{20}{mK}$ at the base stage of a dilution refrigerator. Qubit states are measured using a dispersively-coupled readout resonator~\cite{Krantz2019}, and initial qubit states are heralded with prior readout pulses~\cite{Ding2023}. We realize the system Hamiltonian, written in matrix form with Pauli matrices $\sigma_i$,
\begin{equation}
    \frac{H}{\hbar} = \omega_{01} \frac{\sigma_z}{2} + \eta_x(t)\sigma_x + \eta_y(t)\sigma_y.
\end{equation}
The transverse driving terms originate from charge ($\hat{n} \propto \sigma_y$) and flux ($\hat{\phi} \propto \sigma_x$) control. We include further details on the sample and experimental setup in Supplemental material Section I~\cite{supp}.


The primary experiment performed in this work is a Ramsey experiment with injected noise [\fref{fig1}\panel{a,b}]. Starting from the ground state, we prepare the qubit on the Bloch-sphere equator with a resonant $\pi/2$ pulse. We then let the qubit evolve for a total free-precession time $\tau = 2\tau_b + \tau_n$, where $\tau_n$ is the injected-noise duration and $\tau_b \geq \SI{5}{ns}$ is a buffer time before and after the noise injection, used to minimize interference with the coherent control pulses. Unless specified, we set $\tau_b = \SI{10}{ns}$. The noise is turned on diabatically with rise and fall times of \SI{0.5}{ns} such that the initial qubit state is approximately unchanged when turning the noise on and off. Finally, the qubit state is transferred to the $\hat{z}$ axis with a second resonant $\pi/2$ pulse. We ensure that the initial and final $\pi/2$ pulses are about the same axis in the frame co-rotating with the qubit such that the Bloch vector is aligned with the $\hat{z}$ axis at the end of the sequence. This enables us to probe the purity $\gamma \equiv \text{tr}\left(\rho^2\right)$ with only one measurement of the qubit along $\sigma_z$ as
\begin{equation}
    \gamma \rightarrow \gamma_\text{approx} \equiv \frac{\langle \sigma_z \rangle^2 }{2} + \frac{1}{2}
\end{equation}
in the limit $\langle \sigma_{x} \rangle^2 + \langle \sigma_{y} \rangle^2 \rightarrow 0$ (we justify the correspondence of $\gamma_\text{approx}$ and $\gamma$ in Supplemental Material Section II~\cite{supp}). 
Noise injection is performed by gating a noise source with a double-balanced mixer and connecting it to the charge and/or flux drive line of the qubit [\fref{fig1}\panel{c}]. We begin data collection after a minimum noise duration of approximately $\SI{20}{ns}$ in order to mitigate transients from the rapid toggling of the noise field.

To demonstrate the appearance of purity oscillations from anisotropic noise, we performed Ramsey experiments with noise injected along $\sigma_y$. We utilized three noise configurations: (1) noise off, (2) low-frequency white noise up to $\SI{100}{MHz} < f_{01}$, and (3) broadband white noise up to $\SI{1}{GHz} > f_{01}$ [\fref{fig1}\panel{d}] (see Supplemental Material Section IV for further details about the noise sources~\cite{supp}). We plot the purity decays in \fref{fig1}\panel{e}. With noise off, we observe nearly constant purity due to the long coherence of our qubit relative to the measured free-precession times. With white noise below the qubit frequency, we observe oscillations in the purity at a frequency $2f_{01}$. With white noise over a range that includes the qubit frequency, we observe the purity decay in small steps. We observe oscillations present in the derivative of the signal in \fref{fig1}\panel{e}, inset. For this high-frequency noise, the lack of purity revivals ($\partial \gamma / \partial t \leq 0$) is consistent with the short correlation time of the broadband noise relative to the qubit dynamics, which emulates a Markovian bath. We derive an analytical expression for such purity decays with the Lindblad master equation in Supplemental Material Section VI~\cite{supp}.

\begin{figure}[!htb]
\includegraphics[width=0.45\textwidth]{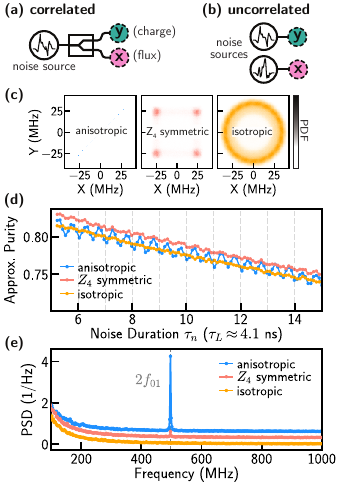}
\caption{
\label{fig2}
\textbf{Varying the anisotropy of injected noise.}
All experiments in this figure utilized the low-frequency white noise source (noise power up to $\SI{100}{MHz} < f_{01}$) with a bimodal amplitude distribution [\fref{fig1}{d}].
\cpanel{a} Correlated (one source split to $\sigma_x$ and $\sigma_y$) and \cpanel{b} uncorrelated (separate sources for $\sigma_x$ and $\sigma_y$) noise source configurations. 
\cpanel{c} Noise-amplitude distributions from noise traces measured on an oscilloscope, reflecting noise in the Bloch-sphere XY plane, for three configurations: (1) anisotropic, comprising the correlated configuration in \panel{a}, (2) $Z_4$-symmetric, comprising the uncorrelated configuration in \panel{b}, and (3) isotropic, comprising the $Z_4$-symmetric noise averaged over 19 equally-spaced rotations about the $\hat{z}$ axis.
\cpanel{d} Time-domain traces of the approximate purity, \approxpurity, during a Ramsey sequence with the injected noise configurations in \panel{c}.  
In order to visualize the oscillations, we plot a subset of the full data (1001 points between $5.12\tau_L \leq \tau_n \leq 105.12\tau_L$).
\cpanel{e} Power spectral densities (PSDs) of the full time-domain data taken for \panel{d}, offset for clarity.
We attribute the small $2f_{01}$ feature in the $Z_4$-symmetric trace to a slight imbalance of the calibrated charge and flux noise amplitudes.
}
\end{figure}

We then validated that the purity oscillations are related to noise anisotropy by measuring the purity dynamics while varying the noise distribution anisotropy between three configurations: (1) anisotropic, (2) $Z_4$-symmetric (symmetric under $90^\circ$ rotations), and (3) isotropic. Anisotropic noise (1) comprised perfectly correlated noise along the $\hat{x}$ (flux) and $\hat{y}$ (charge) axes [\fref{fig2}\panel{a}], yielding noise along $\hat{x} + \hat{y}$. $Z_4$-symmetric noise (2) comprised uncorrelated, equal-amplitude noise sources connected to $\hat{x}$ and $\hat{y}$ [\fref{fig2}\panel{b}]. The $Z_4$ symmetry resulted from the bimodal nature of the noise amplitude distribution (uncorrelated Gaussian noise sources would instead produce isotropic noise). Isotropic noise (3) comprised $Z_4$-symmetric noise averaged over 19 qubit initial states equally spaced along the Bloch-sphere equator, yielding an effective isotropic noise distribution in the XY plane for the ensemble-averaged dynamics~\cite{Gneiting2020}. Two-dimensional noise amplitude distributions are presented in \fref{fig2}\panel{c}. With the anisotropic noise, we observe pronounced oscillations in the time-domain purity trace [\fref{fig2}\panel{d}]. We observe nearly complete extinction of the oscillations with the $Z_4$-symmetric noise and complete extinction for the isotropic noise [\fref{fig2}\panel{e}]. We note that for the isotropic noise, the isotropy is a consequence of ensemble averaging over qubit-state preparations equally spaced along the Bloch-sphere equator~\cite{Gneiting2020} rather than reflecting the noise symmetry during individual shots. We emphasize that the qualitative purity-decay behavior (smooth decay with no oscillations) matches that from numerical simulations with shot-to-shot isotropically distributed noise. We attribute the remaining oscillatory component in the $Z_4$-symmetric trace to imperfect balancing of the noise amplitudes, which were experimentally calibrated by measuring the charge/flux $\pi$-pulse amplitudes individually.

\begin{figure*}[!tb]
\includegraphics[width=\textwidth]{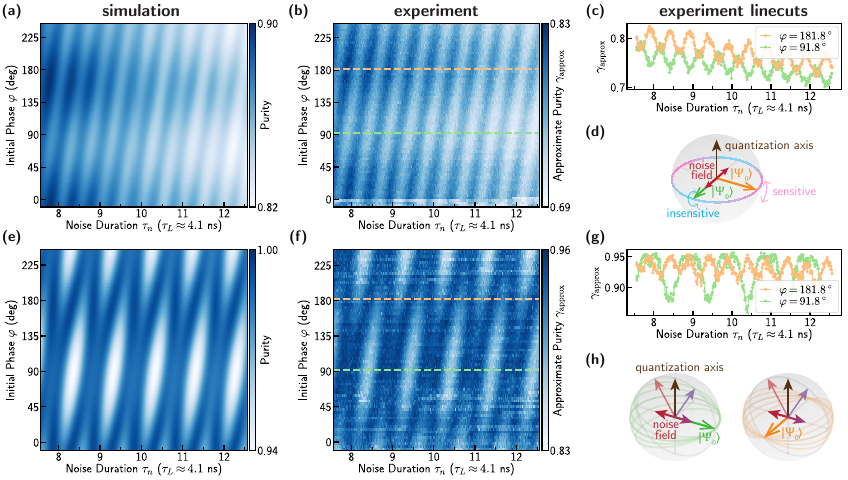}
\caption{
\label{fig3}
\textbf{Noise axis and power spectrum dependence of purity oscillations.}
All data comprise Ramsey experiments with injected charge noise  ($\hat{n} \propto \sigma_y$) as a function of noise duration and initial superposition phase $\varphi$, $\ket{\Psi_0} = (\ket{0} + e^{i \varphi}\ket{1}) / \sqrt{2}$, with $\tau_b = \SI{5}{ns}$.
We elected to sweep the phase $\varphi$ rather than the noise axis for experimental ease, noting that the resulting physics is identical.
Simulated $\SI{100}{MHz}$ white ($\SI{2}{MHz}$ low-pass filtered) noise comprises a random telegraph signal with amplitude $\approx \pm \SI{24}{MHz}$ ($\approx \pm \SI{23}{MHz}$) and average switching rate $\approx \SI{96}{MHz}$ ($\approx \SI{4.4}{MHz}$).
\cpanel{a} Simulation, \cpanel{b} experiment, and \cpanel{c} experimental linecuts for $\SI{100}{MHz}$ white noise. 
\cpanel{d} Bloch sphere cartoon illustrating the insensitivity (sensitivity) of the qubit state to rotations induced by the noise when the qubit is parallel (perpendicular) to the noise axis, resulting in purity oscillations at frequency $2f_{01}$.
\cpanel{e,f,g} Similar to \panel{a,b,c}, respectively, for $\SI{2}{MHz}$ low-pass-filtered noise, which approaches the quasistatic limit where noise is constant during a single run of the experiment (shot), but changes shot-to-shot.
\cpanel{h} Bloch sphere cartoon for quasistatic noise. 
In the quasistatic limit, when noise is constant during the Larmor precession but changing shot-to-shot, the effective quantization axis is affected. 
Left (right): When the qubit state is initially aligned with (perpendicular to) the noise axis, all Larmor orbits have one (two) fixed point(s). This results in maximal purity when the qubit is aligned (aligned or anti-aligned) with its starting state, characterized by purity oscillations of frequency $f_{01}$ ($2f_{01}$).
}
\end{figure*}

As a final main result, we probed the noise-axis dependence of purity oscillations for two anisotropic noise configurations: (1) white noise up to \SI{100}{MHz} [\fref{fig3}\panel{a,b,c}], and (2) \SI{2}{MHz} low-pass-filtered noise [\fref{fig3}\panel{e,f,g}]. We elected to sweep the initial superposition phase $\varphi$ rather than the noise axis directly, which allowed us to calibrate the qubit frequency with the noise-induced AC Stark shift only once at the beginning of the experiment. For noise configuration (1), we observed purity oscillations at twice the qubit frequency with a phase periodicity of $\Delta \varphi = 180^\circ$, consistent with identical purity decays for a shift in the noise axis of $180^\circ$. We can understand the oscillation frequency in the Bloch-sphere picture: as the qubit precesses along the equator, the noise susceptibility oscillates with a period of $\tau_L/2$ [\fref{fig3}\panel{d}]. For noise configuration (2), the correlation time of the noise $\tau_c \sim \SI{100}{ns}$ exceeds the free-precession time, approaching the quasistatic noise limit in which noise is constant during a single experiment, but differs shot-to-shot. We observed purity oscillations with a frequency dependent on the noise axis relative to the initial state, with the same phase periodicity of $\Delta \varphi = 180^\circ$. We can understand the doubling of the period in the Bloch-sphere picture by considering shot-to-shot fluctuations of the qubit quantization axis [\fref{fig3}\panel{h}], and we analytically derive an expression for noise with variance $\langle \eta^2 \rangle$ (see Supplemental Material Section V for details~\cite{supp}):
\begin{equation}\label{eq:quasistatic_ramsey}
    \text{tr}\left(\rho^2\right) = 1 - \frac{2\langle \eta^2 \rangle}{\omega^2}(\sin\varphi + \sin(\omega t - \varphi))^2 + \mathcal{O}\left(\frac{\langle\eta^4\rangle}{\omega^4}\right).
\end{equation}
The purity decay for isotropic quasistatic noise is obtained by averaging \eref{eq:quasistatic_ramsey} over $\varphi \in [0,2\pi)$, yielding purity oscillations at the qubit frequency $\omega$. Unlike the oscillations at $2\omega$, these oscillations are not a consequence of the noise anisotropy; starting in a consistent state on the zero-noise Bloch-sphere equator, tilting the quantization axis in any direction will always leave the starting state as a fixed point of the Larmor orbit~\cite{Gneiting2016}.

We include further experimental and analytical results for relaxation-type experiments in Supplemental Material Section VII~\cite{supp}, where the qubit is prepared in $\ket{1}$ and then subject to anisotropic noise. We observed purity oscillations at the qubit frequency for noise in the quasistatic limit, but not in the Markovian limit. We understand these oscillations as arising from shot-to-shot tilting of the quantization axis, similar to the case described at the end of the previous paragraph.

We now discuss our results in the context of quantum information applications, where understanding and mitigating noise~\cite{Rower2023, Paladino2014, MULLER2019} and further exploring dynamics beyond the RWA~\cite{vacchini_advances_2019, Rower2024, Sank2024} are of high interest due to, e.g., advancing metrology and quantum error correction experiments. For superconducting qubits in particular, decoherence mechanisms are associated with microscopic models that couple to either the charge or flux degrees of freedom of a circuit~\cite{Krantz2019}. 
The time-domain signatures explored in this work can potentially be used to determine if decoherence is due to noise that is anisotropic (i.e., primarily coupled to either charge or flux), and whether the noise dynamics approach either the Markovian or quasistatic limits. 
In this experiment, utilizing a low-frequency fluxonium qubit, no purity oscillations were observed in the case of no injected noise; this may be due to the weakness of the native noise resulting in an oscillation too small to detect, or the low-frequency nature of the qubit which results in decoherence from both $1/f$ flux noise and dielectric loss~\cite{Sun2023, Ateshian2024}. 
For high-frequency qubits such as conventional transmons or fluxonium qubits operated far from the degeneracy point, the dominant transverse loss mechanism may be anisotropic due to, e.g., dominance of the charge or phase matrix element~\cite{Koch2007, Sun2023}, 
or different quantum-to-classical crossover frequencies of charge- and flux-noise spectra~\cite{Astafiev2004, Quintana2017}. 
However, finer time resolution would be required to observe the oscillations due to shorter qubit Larmor periods. 
In general, the signatures explored in this work may be used to validate the understanding of qubit decoherence in other platforms. One potential difficulty of observing such signatures from the native coherence-limiting noise is state preparation: if the noise is present during state preparation, differing initial states shot-to-shot may obfuscate oscillations. For superconducting qubits, this difficulty may be circumvented by preparing initial states at at a low-noise point (e.g., where the charge or flux matrix element is suppressed), and leveraging fast-flux control to probe decoherence where transverse noise is expected to be anisotropic. 

In summary, we investigated qubit decoherence under the effect of transverse noise that is anisotropic (linearly polarized) in the lab frame. We have established that, when initially prepared in a superposition state, the state purity $\text{tr}\left(\rho^2\right)$ displays an oscillatory component at twice the qubit frequency~\cite{vacchini_advances_2019}; when the qubit is aligned with (perpendicular to) the noise, the qubit state is insensitive (sensitive) to noise-induced rotations. We verified that the oscillations are mitigated for noise that is isotropically distributed in the Bloch sphere XY plane and explored the oscillation dependence on the noise axis and power spectral density. We further elucidated our results with analytical models for purity decay in the presence of noise in the quasistatic and Markovian limits. Our results establish the impact of transverse noise anisotropy on qubit decoherence, providing a new time-domain signature for probing the anisotropy of coherence-limiting noise in experiments.

\begin{acknowledgments}
We gratefully acknowledge Prof. Yasunobu Nakamura for useful discussions and support during this project, and Prof. Lorenza Viola, Patrick Harrington, Shoumik Chowdhury, Shantanu Jha, Om Joshi, and Thomas Hazard for insightful conversations.
This research was funded in part by the U.S. Army Research Office under Award No. W911NF-23-1-0045; in part by the U.S. Department of Energy, Office of Science, National Quantum Information Science Research Centers, Co-design Center for Quantum Advantage (C2QA) under contract number DE-SC0012704; and in part under Air Force Contract No. FA8702-15-D-0001.
J.A. greatfully acknowledges the support from the Korea Foundation for Advanced Studies.
M.H. and I.T.R. are supported by an appointment to the Intelligence Community Postdoctoral Research Fellowship Program at MIT administered by Oak Ridge Institute for Science and Education (ORISE) through an interagency agreement between the U.S. Department of Energy and the Office of the Director of National Intelligence (ODNI).
D.A.R. gratefully acknowledges support from the NSF (award DMR-1747426).
S.C.A-F. is supported by the U.S. DOE Office of High Energy Physics under grant number DE-SC0012567.
Any opinions, findings, conclusions or recommendations expressed in this material are those of the author(s) and do not necessarily reflect the views of the US Air Force or the US Government.
\end{acknowledgments}

\bibliography{refs}



\newcommand{\aref}[1]{Supplemental Material Section~\ref{#1}}

\onecolumngrid

\begin{center}
    \textbf{SUPPLEMENTAL MATERIAL}
\end{center}

\setcounter{figure}{0}
\setcounter{equation}{0}
\makeatletter 
\renewcommand{\thefigure}{S\@arabic\c@figure}
\renewcommand{\thetable}{S\@arabic\c@table}
\renewcommand{\theequation}{S\arabic{equation}}

\section{Sample and Experimental Setup}\label{appendix:wiring}
The fluxonium qubit with individual charge and flux control obeys the system Hamiltonian
\begin{equation} \label{eq:full_hamiltonian} 
    \hat H = \hat{H}_0 + \hbar \Omega_{\hat{n}}(t)\hat{n} + \hbar \Omega_{\hat{\phi}}(t) \hat{\phi},
\end{equation}
where $\Omega_{\hat{n}}(t)$ and $\Omega_{\hat{\phi}}(t)$ are the charge and flux drives respectively (comprising both coherent control pulses and injected noise), and $\hat{H}_0$ is the bare fluxonium Hamiltonian 
\begin{equation} 
    \hat H_0 = 4 E_C \hat{n}^2 + \frac{1}{2} E_L (\hat{\phi} - \phi_\text{dc})^2 - E_J \cos(\hat\phi).
\end{equation}
In the fluxonium Hamiltonian, $\hat{n}$ and $\hat{\phi}$ represent the charge and phase operators; $E_C/h = \SI{1.30}{GHz}$, $E_L/h=\SI{0.59}{GHz}$, and $E_J/h=\SI{5.71}{GHz}$ are the charging, inductive, and Josephson energies, respectively; and $\phi_\text{dc}$ is a phase offset resulting from a static external magnetic flux $\Phi_\text{dc} / \Phi_0 = \phi_\text{dc} / 2\pi$ supplied by a superconducting coil mounted to the lid of the sample package, inductively coupled to the fluxonium loop. All experiments were performed with $\Phi_\text{dc} = 0.5\Phi_0$, referred to as the degeneracy point.
The sample qubit was a subsystem of a device comprising two fluxonium qubits with a capacitively-coupled transmon coupler (refer to device A, fluxonium 2 of Ref.~\cite{Ding2023}).

\begin{figure}[!htb]
\includegraphics{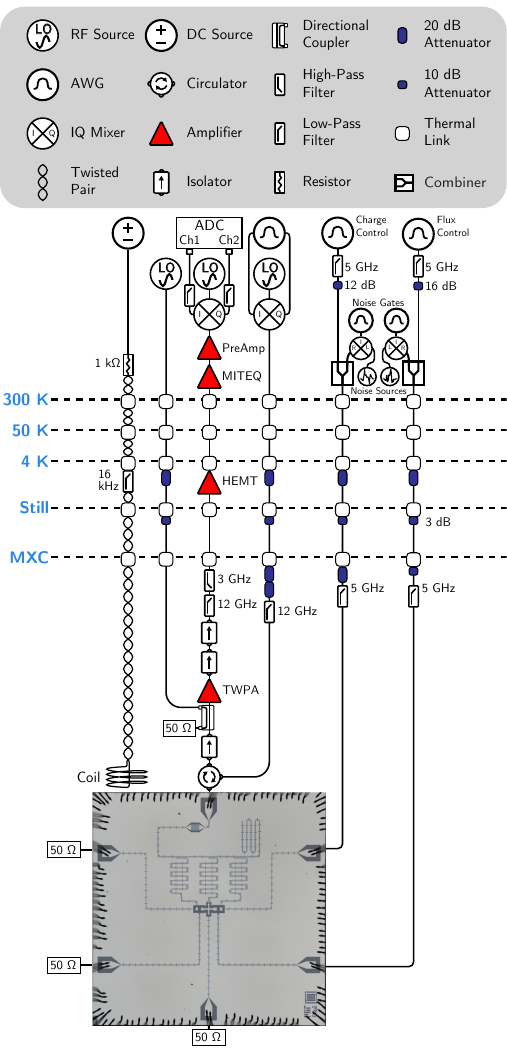}
\caption{\label{fig:sup_wiring} \textbf{Wiring schematic of the experimental setup.}} 
\end{figure}

All experiments were conducted in a Bluefors XLD600 dilution refrigerator maintaining a base temperature stabilized at $\sim \SI{22}{mK}$.
We specify the equipment used for qubit biasing, coherent control pulses, injected noise generation, and readout in \tref{tab:equipment}, and we detail the experimental wiring in \fref{fig:sup_wiring}.

\begin{table}[!htb]
\caption{\label{tab:equipment}
\textbf{Summary of control equipment.}
}
\begin{tabular}{p{6cm} p{5cm}}
\hline
\hline
Component & Manufacturer + Model\\
\hline
Dilution Refrigerator & Bluefors XLD600\\
RF Source & Rohde \& Schwarz SGS100A\\
DC Source & QDevil QDAC I\\
Control Chassis & Keysight M9019A\\
AWG (readout pulses) & Keysight M3202A\\
AWG (qubit pulses) & Keysight M8195A\\
AWG (noise gating) & Keysight M8195A\\
ADC & Keysight M3102A\\
$\SI{100}{MHz}$ white noise & Agilent 33250A\\
$\SI{1}{GHz}$ white noise & (Mini-Circuits ZFL-500LN+)x3\\
Noise $\SI{2}{MHz}$ LPF & Mini-Circuits SLP-1.9+\\
Noise Mixer & Mini-Circuits ZFM-2-S+ \\
Noise Combiner & Mini-Circuits ZFRSC-42-S+ \\
\hline
\hline
\end{tabular}
\end{table}

\section{Approximate Purity}\label{app:approx_purity}
In order to maintain high stability of the readout during measurements, we reduce measurement overhead by approximating the state purity, 
\begin{equation}
    \gamma \equiv  \text{tr}\left(\rho^2\right) = \frac{\langle\sigma_x \rangle^2 + \langle \sigma_y \rangle^2 + \langle \sigma_z \rangle^2}{2} + \frac{1}{2}
\end{equation}
where $\rho$ is the state density matrix, as 
\begin{equation}
    \gamma_\text{approx} \equiv \frac{\langle \sigma_z \rangle^2}{2} + \frac{1}{2}.
\end{equation}
Here, we justify the correspondence of approximate and exact purity.

\begin{figure}[!tb]
\includegraphics[width=0.5\textwidth]{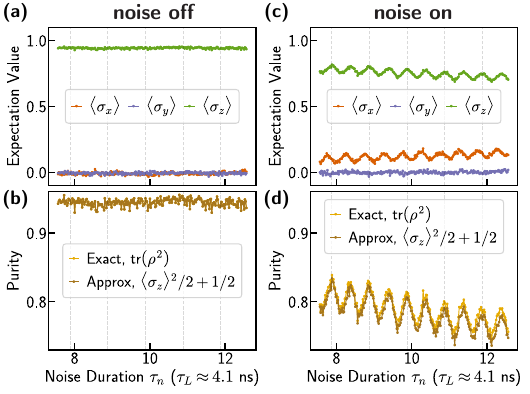}
\caption{
\label{fig:purity}
\textbf{Approximate and exact purity.}
All data shown are for free-induction-decay experiments with injected $\SI{100}{MHz}$ white noise along $\hat{n} \propto \sigma_y$ and buffer time $\tau_b = \SI{5}{ns}$.
\cpanel{a} State tomography and extracted \cpanel{b} exact purity, \purityeq, and approximate purity, \approxpurity, with noise amplitude set to zero.
\cpanel{c,d} Similar to \panel{a,b}, for non-zero noise amplitude.
}
\end{figure}

First, we experimentally confirmed the correspondence of $\gamma$ and $\gamma_\text{approx}$ by performing state tomography after Ramsey experiments with and without injected charge noise of the highest amplitude used in this work [\fref{fig:purity}]. We found that $|\gamma - \gamma_\text{approx}| \leq 1.5\%$ (relative error bounded by $2.1\%$) for free precession times up to $\SI{12.6}{\tau_L}$, validating that $\gamma_\text{approx}$ can act as a reliable proxy for $\gamma$.

We further justify the use of $\gamma_\text{approx}$ analytically. In an ideal Ramsey experiment with the qubit initially in its ground state, the first $\pi/2$ pulse (about the $\hat{y}$ axis, without loss of generality) transfers the qubit state to the $\hat{x}$ axis of the Bloch sphere. 
Subsequently, the Bloch vector rotates about the $\hat{z}$ axis as the qubit undergoes Larmor precession in the laboratory frame. 
In the frame co-rotating with the qubit, the qubit remains stationary and along the $\hat{x}$ axis.
After waiting a set phase accumulation time, a second $\pi/2$ pulse (about the same rotating-frame axis as the first) transfers the qubit state to the $\hat{z}$ axis. 
Assuming that the qubit maintains constant frequency during the phase accumulation time, the length of the Bloch vector at the end of the sequence is given entirely by the expectation value $\langle \sigma_z \rangle$: 
\begin{equation}
    \gamma = \frac{\langle \sigma_z \rangle^2}{2} + \frac{1}{2} = \gamma_\text{approx}.
\end{equation} 
If the qubit is instead subject to a frequency shift during the evolution time, the Bloch vector immediately before the final $\pi/2$ pulse will be slightly misaligned from the $\hat{x}$ axis by an angle $\theta$.
In this case, the approximate purity deviates from the exact purity: 
\begin{equation}
    |\gamma- \gamma_\text{approx}| = \frac{r^2 \text{sin}^2(\theta)}{2},
\end{equation}
where $0 \leq r \leq 1$ is the length of the Bloch vector.

We measured the qubit frequency shift from injected charge noise of the highest amplitude used in this experiment, finding $|f_{01}^\text{bare} - f_{01}^\text{noise}|\lesssim\SI{0.54}{MHz}$. For such noise (as used in Fig. 1, Fig. 3, \fref{fig:purity}, and \fref{fig:T1}) with durations up to $\tau_n \lesssim 12.56 \tau_L$, we expect misalignments from the frequency shift of $\theta \lesssim 10^\circ$ and resulting errors in the approximate purity bounded by $|\gamma - \gamma_\text{approx}| \leq 1.5\%$, consistent with measured data.

We found that injected flux noise significantly modified the qubit frequency. For the configuration used in Fig. 2 with simultaneously injected charge and flux noise, we measured a frequency shift of $|f_{01}^\text{bare} - f_{01}^\text{noise}|\approx 4.6 \pm \SI{0.015}{MHz}$. Accordingly, we adjusted the final $\pi/2$-pulse phase for the experiments in Fig. 2 to account for the measured detuning. The frequency uncertainty from the fit leads to an error bound of the approximate purity of $< 10^{-4}$ for noise durations up to $15\tau_L$. 

\section{Frequency Dependence of Purity Oscillations}

\begin{figure}[!tb]
\includegraphics[width=0.5\textwidth]{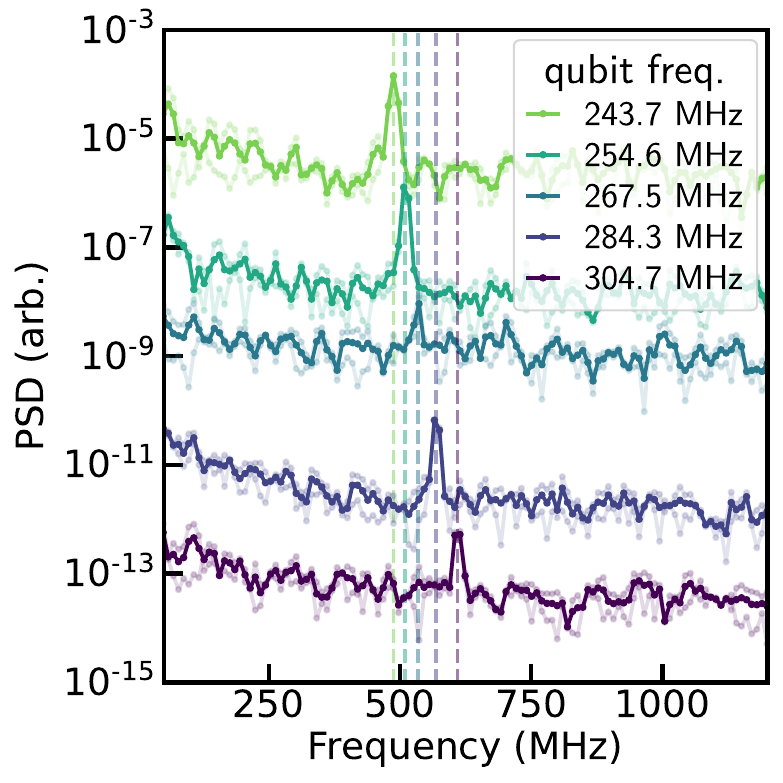}
\caption{
\label{fig:freq}
\textbf{Purity oscillations vs. qubit frequency.}
Each trace corresponds to a different flux bias, and represents a PSD calculated with Welch's method for a free-induction decay experiment with injected $\SI{100}{MHz}$ white noise along $\hat{n} \propto \sigma_y$ and $\tau_b = \SI{5}{ns}$.
Traces are artificially offset for clarity.
Dashed lines correspond to $2f_{01}$.
}
\end{figure}

In order to validate that the frequency of the purity oscillations was determined by the twice the qubit frequency, we measured purity oscillations from injected charge noise as a function of external flux bias in a small window around $\Phi_\text{ext} = 0.5 \Phi_0$ corresponding to qubit frequencies $\SI{243.7}{MHz} \leq f_{01} \leq \SI{304.7}{MHz}$ [\fref{fig:freq}]. At each external flux bias, readout pointers, the qubit frequency, and the $\pi/2$-pulse amplitude was recalibrated. We observed the purity oscillation frequency matched $2f_{01}$ at all measured biases, consistent with the oscillations being generated by the anisotropic noise.

\section{Injected Noise Characterization and Simulation}\label{supp:psd}
In this section, we present characterization of the injected noise used in our experiments and the simulation methodology used to emulate the injected noise. Before detailing the particular noises used, we emphasize that at the level of qualitative behavior (e.g., presence or absence of purity oscillations), all simulated signals did not depend sensitively on the noise distribution (bimodal, Gaussian, or uniform) or on the high-frequency limit of the noise PSD ($1/f^2$ or $1/f^4$), consistent with the analytical results of Supplemental Material, Sections \ref{supp:analysis_quasistatic} and \ref{supp:analysis_lindblad}.

\begin{figure}[!htb]
\includegraphics[width=0.6\textwidth]{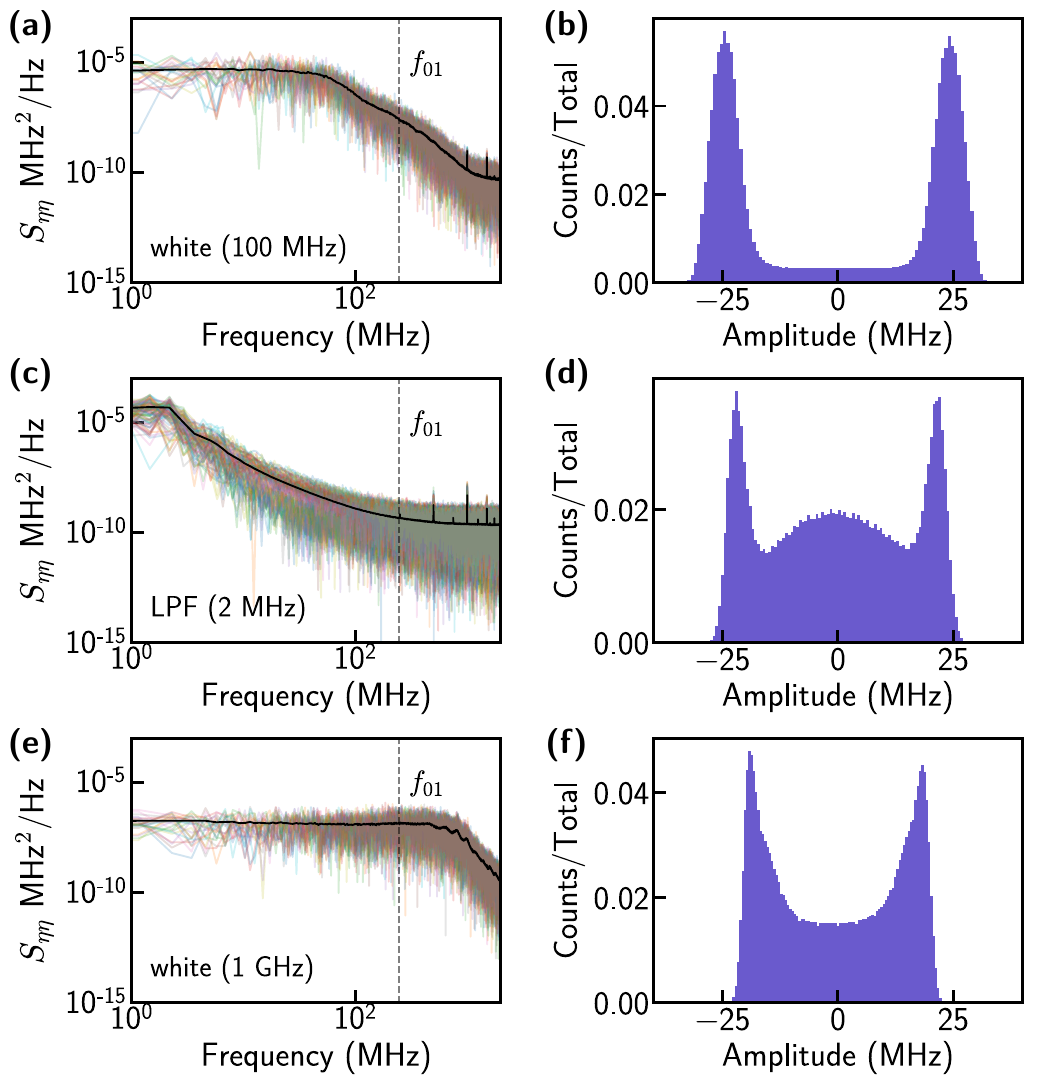}
\caption{
\label{fig:psd}
\textbf{Injected noise characterization.}
To most directly characterize the noise sent to the qubit, all data presented were measured with an oscilloscope after the noise gating and combination with the coherent control line.
\cpanel{a} Noise power-spectral density (PSD) and \cpanel{b} noise amplitude distribution for the $\sim\SI{100}{MHz}$ white noise configuration.
\cpanel{c,d} Similar to \panel{a,b}, for $\SI{2}{MHz}$ low-pass filtered noise.
\cpanel{e,f} Similar to \panel{a,b}, for $\SI{1}{GHz}$ white noise.
}
\end{figure}

To characterize the injected noise, we measured noise after the combiner as in the wiring diagram [Fig. 1\panel{b}] in the time domain with a high-bandwidth oscilloscope for the three noise configurations used in this work: (1) white noise up to $\sim\SI{100}{MHz}$ [\fref{fig:psd}\panel{a,b}] generated with an Agilent 33250A, (2) $\SI{2}{MHz}$ low-pass filtered noise [\fref{fig:psd}\panel{c,d}] generated by filtering the output of the Agilent 33250A before mixing with the noise gate, and (3) white noise up to $\SI{1}{GHz}$ [\fref{fig:psd}\panel{e,f}] generated by daisy-chaining three amplifiers with a passive $\SI{50}{Ohm}$ input. For each type of noise, we computed the PSD of measured time-domain traces, and averaged the resulting PSD over several noise instances. The PSD was computed as
\begin{equation}
    S_{\eta\eta}(f) = \frac{|\text{FFT}\{\eta(t)\}|^2 \delta t}{N},
\end{equation}
where $\eta(t)$ is the noise process, sampled at rate $1/\delta t$ yielding a time-series with $N$ total samples. We converted the units of the time series from voltage to frequency by scaling the measured traces by the ratio of the analytical $\pi/2$ amplitude for our $\SI{80}{ns}$ cosine pulse (in MHz) to the measured $\pi/2$ pulse amplitude (in volts). We report the measured PSDs and amplitude distributions in \fref{fig:psd}.

The $\SI{100}{MHz}$ white noise amplitude distribution most closely resembled an ideal bimodal distribution, and the $\SI{2}{MHz}$ low-pass filtered and $\SI{1}{GHz}$ white noise configurations contained an additional significant proportion of the amplitude distribution centered around zero. We attribute the sharp cutoff of the noise amplitude distribution to saturation of the mixer used for gating the noise. We emphasize that in simulations, we found no dependence of the presence of purity oscillations on the specific noise amplitude distribution (e.g., bimodal, Gaussian, or uniform) for a given noise power spectral density. Our analytical treatment of purity decays suggests that higher-order moments of the noise distribution would add small effects on the order of $(\eta / \omega)^n\sim(10\%)^n$ for $n>2$. 

\begin{figure}[!htb]
\includegraphics[width=0.45\textwidth]{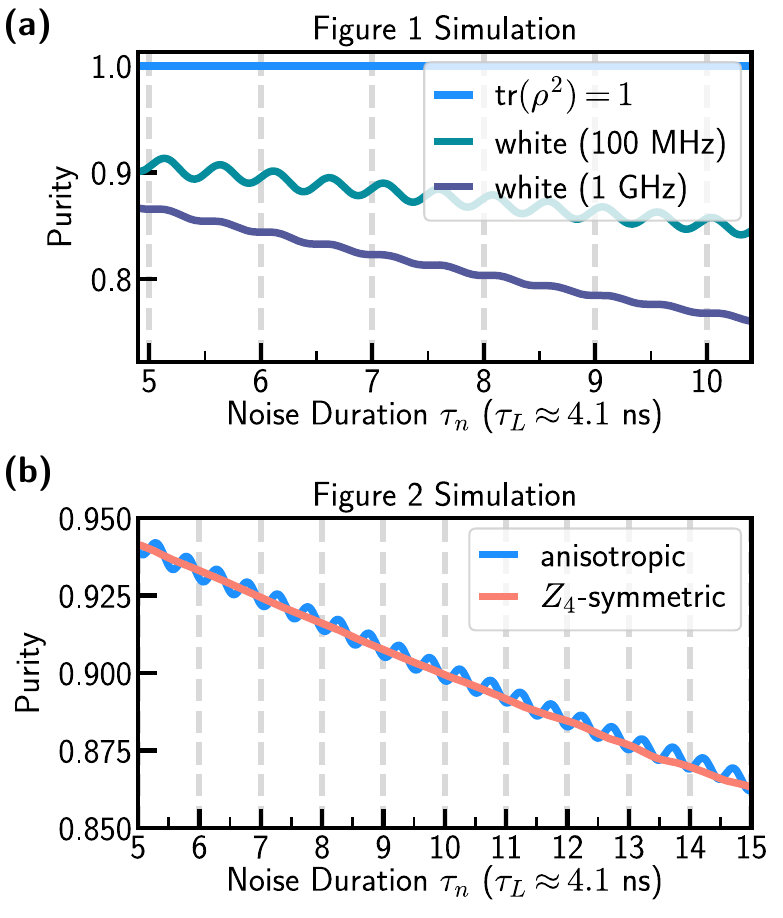}
\caption{
\label{fig:sim}
\textbf{Simulation of Figures 1 \& 2.}
Simulations were performed with the methodology described in \aref{supp:psd}.
\cpanel{a} Simulated Ramsey experiments as in Fig. 1, with injected $\sigma_y$ noise in three configurations: (1) no noise, represented by a pure state with $\text{tr}\left(\rho^2\right) = 1$ (light blue), (2) low-frequency white noise up to $\SI{100}{MHz} < f_{01}$ (turquoise), and (3) broadband white noise up to $\SI{1}{GHz} > f_{01}$ (dark blue).
\cpanel{b} Simulated Ramsey experiments as in Fig. 2, with $Z_4$-symmetric noise (uncorrelated equal-amplitude noise along $\sigma_x$ and $\sigma_y$, red), and anisotropic noise (correlated equal-amplitude noise along $\sigma_x$ and $\sigma_y$, blue).
}
\end{figure}

To simulate our experimental results, we first extracted the correlation time of the measured injected noise for each configuration. We then modelled the noise as an ideal random telegraph signal (RTS) with average switching rate determined by the correlation time, and amplitude determined by the cutoff values (defined as the center of the peaks at the edges of the distributions in \fref{fig:psd}). For each simulated experiment, we generated several instances of noise $\eta(t)$ and calculated the noise-averaged purity as $\text{tr}\left(\rho^2\right)$, where $\rho = \langle \rho_{\eta(t)} \rangle$ and $\langle \cdot \rangle$ represents the noise-ensemble average. We plot simulated experiments corresponding to main text Figures 1 \& 2 in \fref{fig:sim}. We find agreement of the qualitative behavior between simulation and experiment; anisotropic noise leads to purity oscillations at twice the qubit frequency, with revivals (monotonic decay) generated by noise with a long (short) correlation time [\fref{fig:sim}\panel{a}], and extinction of the oscillations for noise which is isotropically distributed in the Bloch-sphere XY plane [\fref{fig:sim}\panel{b}]. We find quantitative order-of-magnitude agreement of the oscillation fringe contrast and decay magnitudes between simulation and experiment. We note that differences can arise from a variety of sources including state preparation and measurement (SPAM) errors, differences between the simulated (ideal RTS) and experimentally realized noise, and discrepancies between the approximate purity (reported for experiments) and true purity (reported for simulations). 

\section{Analytical Results for Anisotropic Quasistatic Noise}\label{supp:analysis_quasistatic}
In this section, we derive expressions for the purity $\gamma = \text{tr}\left(\rho^2\right)$ of a qubit prepared in a superposition state $\ket{\Psi_0} = (\ket{0} + \ket{1})/\sqrt{2}$ and evolved under the influence of a static noise field $\eta \in \mathbb{R}$ coupled to an axis at angle $\theta$ with the Bloch-sphere $\hat{x}$ axis. The Hamiltonian is given by
\begin{equation}\label{eq:ham_arb_axis_noise}
    \frac{H_\eta}{\hbar} = -\omega \frac{\sigma_z}{2} + \eta (\cos(\theta)\sigma_x + \sin(\theta)\sigma_y).
\end{equation}
Here, $\eta \ll \omega$ is a random variable representing noise that is constant during a single experiment but different shot-to-shot. We refer to such noise as ``quasistatic.'' For a single instance of noise $\eta$, the qubit state at time $t$ will be given by $\ket{\Psi_\eta(t)} = U_\eta(t) \ket{\Psi_0}$, where $U_\eta(t)$ is the time-evolution operator for $H_\eta$.

We compute the noise-averaged purity from the density matrix $\rho_\eta = \ket{\Psi_\eta}\bra{\Psi_\eta}$ as $\rho = \langle \rho_\eta \rangle$, where $\langle \cdot \rangle$ represents the noise ensemble average~\cite{Gneiting2020}. In the frame co-rotating with the qubit and keeping only terms up to $(\eta/\omega)^2$, we find 
\begin{align}
    \tilde{\rho}^{00}_\eta &= \frac{1}{2} + \frac{\eta}{\omega} \left( \cos(\theta + \omega t) - \cos(\theta) \right) \\
    \tilde{\rho}^{11}_\eta &= \frac{1}{2} - \frac{\eta}{\omega} \left( \cos(\theta + \omega t) - \cos(\theta) \right) \\
    \tilde{\rho}^{01}_\eta &= \frac{1}{2} - \frac{\eta^2}{\omega^2}\left[ 1 - i \omega t - e^{-i(2\theta + \omega t)}(1 + e^{2i\theta} - \cos(\omega t))\right] \\
    \tilde{\rho}^{10}_\eta &= (\tilde{\rho}^{01}_\eta)^*,
\end{align}
and note that $\text{tr}(\tilde{\rho}_\eta^2) = \text{tr}(\rho_\eta^2) = 1$ up to terms of order $(\eta/\omega)^4$.  For zero-mean noise processes ($\langle \eta \rangle = 0$, as was the case for all experiments and simulations in this work), we find
\begin{equation}
    \label{eq:quasistatic_purity}
    \text{tr}\left(\rho^2\right) = 1 - \frac{2\langle \eta^2 \rangle}{\omega^2}(\cos\theta - \cos(\omega t + \theta))^2 + \mathcal{O}(\langle\eta^4\rangle/\omega^4).
\end{equation}

We can understand the period-doubling phenomenon of Fig. 3\panel{e,f,g} by expanding the squared second term, yielding two time-dependent oscilations: (1) $\cos^2(\omega t + \theta)$ and (2) $\cos\theta\cos(\omega t + \theta)$. To directly compare with the data, we note that $\theta = \pi/2 - \varphi$ where $\varphi$ is the initial superposition phase as defined in Fig. 3. When the noise axis is perpendicular to the initial qubit state ($\theta = \pi/2 + \pi n$ for integer $n$), the second oscillation vanishes and the remaining oscillation has a period of $\tau = \pi/\omega = \tau_L/2$. In contrast, when the noise axis is aligned or anti-aligned with the initial qubit state ($\theta=n \pi$), both oscillations remain and the period is determined by the slowest oscillation with $\tau = 2\pi/\omega = \tau_L$. We further note that the purity decay has periodicity in $\theta$ of $\Delta\theta = \pi$, consistent with the data of Fig. 3\panel{e,f}.


\section{Analytical Results for Anisotropic White Noise}
\label{supp:analysis_lindblad}

In this section, we derive expressions for purity decays in the case of transversely-coupled high-frequency white noise by modelling the qubit as an open quantum system for which the noise acts as a Markovian bath.
Notably, the Markovian model for linearly-polarized (maximally anisotropic) white noise leads both to an exponential decay of purity -- characteristic of dissipative dynamics in open quantum systems -- as well as oscillations in purity at twice the qubit frequency.

In experiments, we realize high-frequency white noise with the Hamiltonian
\begin{equation}
    \frac{H_\eta}{\hbar} 
    =
    -\omega \frac{\sigma_z}{2}
    +
    \eta(t) \left(
        \cos(\theta)\sigma_x 
        +
        \sin(\theta)\sigma_y
    \right)
    ,
\end{equation}
where \(\eta(t)\) comprises a noise process, with a white power spectral density extending to frequencies above the qubit frequency $\omega$ [\fref{fig:psd}\panel{e,f}], and \(\theta\) indicates the polarization axis of the noise with respect to the \(\hat{x}\) axis of the Bloch sphere.
In this section, however, we model the evolution of the qubit as Markovian by using a Lindblad master equation for the qubit density matrix \cite{Lindblad:1975ef,Gorini:1975nb},
\begin{align}
    \dot\rho
    =
    -i
    \left[
        \frac{H}{\hbar},
        {\rho}
    \right]
    +
    \sum_i
    \Gamma_i\left(
        L_i \rho L^\dagger_i
        -
        \frac{1}{2}
        \left\{
            L_i^\dagger L_i,
            \rho
        \right\}
    \right)
    ,
\end{align}
where the \(L_i\) are a set of jump operators describing the Markovian dynamics of the noise, and the \(\Gamma_i \geq 0\) is a dissipation rate characterizing the strength of the coupling between the white noise and the qubit.
The qubit's evolution becomes Markovian if the qubit and the noise bath remain roughly unentangled and uncorrelated over the course of the qubit's evolution.
See, for example, Ref.~\cite{Preskill:98}.

For transversely-polarized noise, we use the simple model of a single jump operator
\begin{equation}
    L = \cos(\theta)\sigma_x+\sin(\theta)\sigma_y,
\end{equation}
and a single corresponding dissipation rate \(\Gamma\).
This choice represents a coupling of the qubit to a stochastic noise field along the \(\left(\cos(\theta) \, \hat{x} + \sin(\theta) \, \hat{y}\right)\) axis of the Bloch sphere \cite{vacchini_advances_2019, Kozbial2024}.
The density matrix equation of motion is given by
\begin{equation}\label{sup:lindblad_rhodot}
    \dot{\rho} = \begin{pmatrix}
        \Gamma (\rho_{11}-\rho_{00}) & (i\omega - \Gamma)\rho_{01} + e^{-2i\theta}\Gamma\rho_{10} \\
        -(i\omega + \Gamma)\rho_{10} + e^{2i\theta}\Gamma\rho_{01} & \Gamma (\rho_{00}-\rho_{11}) 
    \end{pmatrix}.
\end{equation}
We can gain some insight by looking at the resulting Bloch equations for such single-axis noise, taking $\theta=0$ without loss of generality,
\begin{align}
    \frac{\partial \langle \sigma_x \rangle}{\partial t} &= \omega \langle \sigma_y \rangle \\ 
    \frac{\partial \langle \sigma_y \rangle}{\partial t} &= -\omega \langle \sigma_x \rangle - 2\Gamma \langle \sigma_y \rangle \\
    \frac{\partial \langle \sigma_z \rangle}{\partial t} &= - 2\Gamma \langle \sigma_z \rangle.
\end{align}
We find relaxation of the transverse polarization solely along $\hat{y}$ and not $\hat{x}$ (see Ref. ~\cite{slichterPrinciplesMagneticResonance1990}, Section 5.12 for an alternative derivation). These equations are identical to those describing the infinite-temperature, zero-bias spin-boson model in the weak-coupling limit, see Ref.~\cite{Leggett1987}, Section IIIB, Eq. 3.9 (with our spin quantized along $\hat{z}$ rather than $\hat{x}$, and noise along $\hat{x}$ rather than $\hat{z}$). We note the well-established result that the transverse polarization dynamics, e.g. $\langle \sigma_x\rangle(t)$, are equivalent to those of a damped harmonic oscillator with bare frequency $\omega$. However, time-domain decoherence signatures of the noise anisotropy at frequency $2\omega$ are uncovered by studying the state purity. 
Starting with a qubit prepared in the state \(\ket{\Psi_0} = \left(\ket{0}+\ket{1}\right)/\sqrt{2}\), we solve \eref{sup:lindblad_rhodot}, finding
\begin{equation}
    \label{eq:lindblad_purity}
    \text{tr}\left(\rho^2\right)
    =
    1 - \Gamma t
    - \frac{\Gamma}{2\omega}
    \left(
        \sin(2\theta) - \sin(2\theta + 2\omega t)
    \right)
    +
    \mathcal{O}(\Gamma^2)
    ,
\end{equation}
which exhibits oscillations at twice the qubit frequency and a periodicity in the noise axis of $\Delta \theta = \pi$.
In contrast to the case of low-frequency noise [Fig. 3\panel{a,b,c}], we find that the purity decay for high-frequency white noise is monotonic: $\partial\gamma/\partial t \leq 0$.  We find that the derivative has oscillation amplitude $0 \leq |\partial \gamma / \partial t| \leq 2\Gamma$. From the experimental data in Fig. 1\panel{e} for broadband white noise, fitting the decay envelope to an exponential function yields $\Gamma_\text{from envelope} = 6.4 \pm \SI{0.5}{MHz}$. We corroborate the analytical result by inspecting the initial fringe contrast of the derivative in Fig. 1\panel{d}\panel{e}, inset, yielding $\Gamma_\text{from derivative} \sim \SI{5}{MHz}$. Due to noisiness of the signal, we only emphasize that both estimates of $\Gamma$ yield order-of-magnitude agreement.

We note that for isotropic (unpolarized) noise, qubit purity decays exponentially with no oscillations.
For example, choosing \(\Gamma_1 = \Gamma_2 = \Gamma\) and \(\left(L_1, L_2\right) = \left(\sigma^+, \,\, \sigma^-\right)\) or \(\left(L_1, L_2\right) = \left(\sigma_x/\sqrt{2}, \, \sigma_y/\sqrt{2}\right)\), we find
\(\text{tr}\left(\rho^2\right) = (1+e^{-2\Gamma t})/2\).
This result matches the qualitative conclusions of Fig. 2\panel{d}. However, we note that the noise used in the isotropic noise experiment did not approach the Markovian limit. In numerical simulations of low-frequency $Z_4$-symmetric noise (between the quasistatic and Markovian limits), we find a similar qualitative result: monotonic purity decay without oscillations or step-wise degradation [\fref{fig:sim}\panel{b}].
%


\section{Relaxation Experiments}\label{supp:relaxation}
\begin{figure}[!htb]
\includegraphics[width=0.5\textwidth]{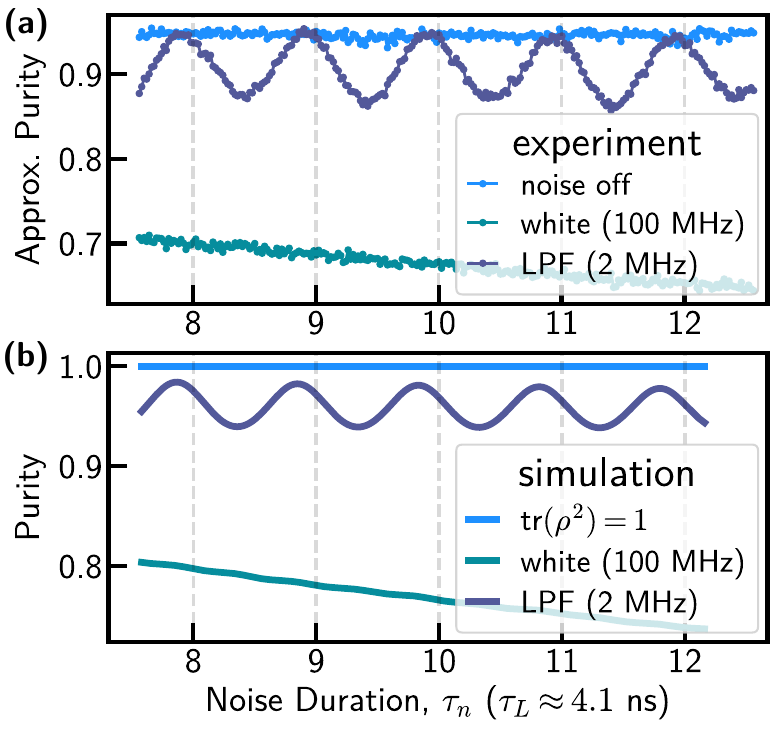}
\caption{
\label{fig:T1}
\textbf{Purity oscillations in relaxation experiments.}
\cpanel{a} Experiment and \cpanel{b} simulation for relaxation experiments with $\ket{\Psi_0} = \ket{1}$ and injected noise along $\sigma_y$, $\tau_b = \SI{5}{ns}$.
Traces are shown for the case of zero noise (light blue), white noise up to $\SI{100}{MHz}$ (green), and $\SI{2}{MHz}$ low-pass filtered noise (dark blue). 
Low-pass filtered noise perturbs the quantization axis of the qubit, leading to purity oscillations at frequency $f_{01}$ from the modified Larmor precession.
}
\end{figure}

In this section, we explore the affect of anisotropic (linearly polarized) noise on relaxation experiments where the qubit is initially prepared in $\ket{\Psi_0} = \ket{1}$ and then subject to noise injected on the charge line ($\hat{n} \propto \sigma_y$): $H_\text{noise} = \eta_y(t) \sigma_y$. We perform relaxation experiments with injected noise in the low-frequency and quasistatic limits [\fref{fig:T1}]. Interestingly, we observe pronounced purity oscillations for experiments with noise approaching the quasistatic limit. Through further numerical simulations and analytical results, we establish that these oscillations do not depend on the axis of the noise and do not depend sensitively on the distribution of noise amplitudes (e.g., Gaussian or bimodal). 

Here we give intuition for these purity oscillations for relaxation experiments with classical transverse noise in the quasistatic limit. Consider a qubit initially prepared in $\ket{1}$, and subject to time-evolution governed by the system Hamiltonian \eref{eq:ham_arb_axis_noise}. The Bloch vector, starting at the pole, will precess around the new effective quantization axis with a period $\omega_\eta = \sqrt{\omega^2 + 4 \eta^2} \approx \omega + 2 \eta^2 / \omega$. For any $\eta$, the Larmor orbit will always pass through the pole of the zero-noise Bloch sphere resulting in maximum purity. Oscillations in the purity will be pronounced until the orbits for different $\eta$ diverge enough to average incoherently. We can estimate the damping time for these oscillations by considering how long it takes for $\pi$ phase difference to accumulate between orbits for noise of strength $\eta$ and no noise: $\Delta T = \pi / |\omega - \omega_\eta| \approx \pi \omega / 2 \eta^2 = f_{01} / 4 (\eta/2\pi)^2$. For the simulated quasistatic noise used in \fref{fig:T1}, we calculate $\Delta T \approx 29 \tau_L$, which is consistent with the $1/e$ decay time of the oscillation amplitude to within $15\%$.

We can understand the oscillations analytically by repeating the analysis for quasistatic noise presented in \aref{supp:analysis_quasistatic}, with the modification that $\ket{\Psi_0} = \ket{1}$. For a noise value $\eta$, the density matrix in the frame co-rotating with the qubit is given by
\begin{align}
    \tilde{\rho}^{00}_\eta &= \frac{2\eta^2}{\omega^2}(1-\cos(\omega t)) \\
    \tilde{\rho}^{11}_\eta &= 1 - \tilde{\rho}^{00}_\eta \\
    \tilde{\rho}^{01}_\eta &= \frac{\eta}{\omega} e^{-i \theta}(-1 + e^{-i \omega t}) \\
    \tilde{\rho}^{10}_\eta &= (\tilde{\rho}^{01}_\eta)^*,
\end{align}
with $\text{tr}(\rho_\eta^2) = 1$ up to terms of order $(\eta/\omega)^4$. For zero-mean noises ($\langle \eta \rangle = 0$), we find
\begin{equation}
    \text{tr}\left(\rho^2\right) = 1 - \frac{4\langle\eta^2\rangle}{\omega^2}(1-\cos(\omega t)) + \mathcal{O}(\langle\eta^4\rangle/\omega^4).
\end{equation}

We note that there is no $\theta$ dependence, i.e., these purity oscillations have no dependence on the noise axis. From the experimental data of \fref{fig:T1}\panel{a} with fringe contrast $\sim 4.3\%$ and $\omega/2\pi \approx \SI{243.7}{MHz}$, we estimate $\sqrt{\langle \eta^2 \rangle} \approx \SI{19}{MHz}$, consistent with the measured injected noise standard deviation of $\sqrt{\langle \eta_\text{meas}^2\rangle}\approx\SI{16}{MHz}$. 

We emphasize that the result for relaxation experiments stands in contrast to our result for Ramsey experiments, for which purity oscillations have a noise-axis-dependent period [Fig. 3\panel{e,f,g}]. We confirmed the lack of noise-axis dependence for relaxation experiments with quasistatic noise in further numerical simulations. We also find that for polarized Markovian noise treated with the Lindblad equation as in \aref{supp:analysis_lindblad} with $\ket{\Psi_0} = \ket{1}$, relaxation experiments result in purely exponential purity decays with no noise axis dependence.

\end{document}